\documentclass[referee,a4paper]{raa}
\usepackage{graphicx,times}
\usepackage{amssymb,amsmath}
\usepackage{geometry}
\newcommand{\mbh}{M_{\rm BH}}
\newcommand{\msun}{{\rm M}_{\odot}}
\newcommand{\ledd}{L_{{\rm Edd}}}
\newcommand{\mdot}{\dot M}
\newcommand{\medd}{{\dot M_{\rm Edd}}}
\newcommand{\ergs}{{\rm \,erg\,s^{-1}}}
\newcommand{\kev}{{\rm\ keV}}

\begin{document}

\title{Jet-dominated quiescent state in black hole X-ray binaries: the cases of A0620--00 and XTE J1118+480%$^*$
%\footnotetext{\small $*$ Supported by the National Natural Science Foundation of China.}
}
\volnopage{ {\bf 2015} Vol.\ {\bf X} No. {\bf XX}, 000--000}
\setcounter{page}{1}
\author{Qi-Xiang Yang\inst{1,2}}
\institute{Key Laboratory for Research in Galaxies and Cosmology, Shanghai Astronomical Observatory, Chinese Academy of Sciences, 80 Nandan Road, Shanghai 200030, China;
 {\it qxyang@shao.ac.cn}\\
\and
University of Chinese Academy of Sciences, 19A Yuquan Road, Beijing 100049, China.\\
\vs \no
   {\small Received **** ; accepted ****}
}

\abstract{The radiative mechanism of the black hole X-ray transients (BHXTs) in their quiescent states (defined as the 2-10$\kev$ X-ray luminosity $\lesssim 10^{34}\ergs$) remains unclear. In this work, we investigate the quasi-simultaneous quiescent state spectrum (including radio, infrared, optical, ultraviolet and X-ray) of two BHXTs, A0620--00 and XTE J1118+480. We find that, these two sources can be well described by a coupled accretion -- jet model. More specifically, most of the emission (radio up-to infrared, and the X-ray waveband) comes from the collimated relativistic jet. Emission from hot accretion flow is totally insignificant, and it can only be observed in mid-infrared (the synchrotron peak). Emission from the outer cold disc is only evident in UV band. These results are consistent with our previous investigation on the quiescent state of V404 Cyg, and confirm that the quiescent state is jet-dominated.
}
\authorrunning{Q. X. Yang}            %author_head in even pages
\titlerunning{Jet-dominated quiescent state in BHXTs}  % title_head in odd pages
   \maketitle

\section{Introduction}           %% first-level sections will be auto-capitalized
\label{intro}

Black hole X-ray transients (BHXTs) are binary systems in which the black hole (BH) accretes matter from its companion. For the majority of time, the BHXTs are observed to be extraordinarily faint, with X-ray (in the energy band 2-10 $\kev$) luminosity $L_{\rm X} \lesssim 10^{34}\ \ergs \lesssim 10^{-5}\ \ledd$.\footnote{The Eddington luminosity is $\ledd = 1.3\times10^{38}\ (\mbh/\msun)\ \ergs$, where $\mbh$ is the black hole mass.} Terminologically, this faint period is called the ``quiescent state''. Such faint accretion phase is also observed in accretion systems around supermassive black holes (SMBHs), i.e. most of the nearby active galactic nuclei (AGN) are systems accreting at low luminosities, with $L_{\rm X} \lesssim 10^{-6}\ \ledd$ (e.g. Ho ~\cite{ho08, ho09}; Pellegrini ~\cite{pel10}). It is further argued that most of the SMBHs remain non-active during their lifetime.

Occasionally, with long intervals (years to decades) staying in quiescent state, BHXTs will undergo outbursts, during which they exhibit distinctive states (soft, hard and intermediate) according to the spectral and timing properties (Zdziarski \& Gierlinski ~\cite{zg04}; Homan \& Belloni ~\cite{hb05}; Remillard \& McClintock ~\cite{rm06}; Done, Gierlinski \& Kubota ~\cite{done07}; Belloni ~\cite{b10}), likely a consequence of the changes of accretion modes (e.g. Esin et al. ~\cite{esin97}). It is now widely accepted that the spectrum of the soft state can be described by a cold multi-temperature blackbody emission, i.e. the accretion flow is a geometrically-thin Shakura-Sunyaev disc (Shakura \& Sunyaev ~\cite{ss73}, hereafter SSD) extending down to the innermost stable circular orbit (ISCO). Hot corona sandwiching the SSD will emit power-law X-ray spectra. The accretion flow and the radiative mechanism of the hard state, on the other hand, is still under active debate. Several models with different dynamics are proposed, i.e. the maximally efficient jet model (Markoff, Nowak \& Wilms ~\cite{m05}; Kylafis et al. ~\cite{k08}), the jet-emitting-disc model (e.g. Ferreira et al. ~\cite{Ferr06}; Zhang \& Xie ~\cite{zx13}), the evaporated-corona model (Liu et al.~\cite{liu02,liu07}; Qiao \& Liu ~\cite{ql13}), and finally our favorite, the accretion -- jet model (Esin et al.~\cite{esin97}; Yuan, Cui \& Narayan ~\cite{ycn05}, hereafter YCN05. See Yuan \& Narayan ~\cite{yn14} for an up-to-date review on this model). More details on the accretion -- jet model will be given later in Sec.\ \ref{sec:accmodel}. Besides, the collimated relativistic jet, which is most evident in radio band, is also observed to be correlated with the accretion mode (e.g. Fender, Belloni \& Gallo ~\cite{f04}),  it is evident in the hard state but will be highly suppressed during the soft state.

Compared to soft and hard states, the nature of the quiescent states in BHXTs remains even more unclear (Narayan, Garcia \& McClintock~\cite{n02}; Narayan \& McClintock~\cite{nm08}; Xie, Yang \& Ma~\cite{x14}, hereafter XYM14; Plotkin et al.~\cite{p15}). Extensive efforts are devoted to understand the quiescent state of BHXTs during the past decade. Observationally, it is well known that the X-ray spectrum of the quiescent state is much softer than hard states, with the photon index $\Gamma$ (defined as the flux at given frequency $F_\nu \propto \nu^{1-\Gamma}$) plateaus to an average $\langle\Gamma\rangle\approx2.1$ (Kong et al~\cite{Kong02}; Corbel, Tomsick \& Kaaret~\cite{c06}; Pszota et al.~\cite{p08}; Reynolds \& Miller~\cite{r11};  Plotkin, Gallo \& Jonker~\cite{p13}; Reynolds et al.~\cite{r14};  Bernardini \& Cackett~\cite{bc14}; Yang et al.~\cite{Yang15}). These observations indicate that the quiescent state may be different from the hard states in its radiative mechanism. Besides, the optical and X-ray variabilities in the quiescent state are tightly correlated (Hynes et al.~\cite{h04}; Reynolds \& Miller~\cite{r11}. See XYM14 for a theoretical interpretation.), e.g. in V404 Cyg. Moreover, recently it is found that the quiescent state of BHXTs is not silent or quiet as expected, but instead shows numerous weak activities (e.g. Cantrell et al. ~\cite{Cantrell10}; Khargharia et al. ~\cite{Khargharia13}; Bernardini \& Cackett ~\cite{bc14}; Rana et al. ~\cite{r15}).

Theoretically, several models are proposed for the X-ray emission in quiescent state. It could be the synchrotron radiation from the non-thermal electrons in the jet (e.g. Yuan \& Cui ~\cite{yc05}; Pszota et al.~\cite{p08}), or the Comptonized emission with seed photons (synchroton) in ADAF (e.g. Narayan, McClintock \& Yi~\cite{n96}; Narayan, Barret \& McClintock~\cite{n97}), or the synchrotron self-Compton (SSC) processes with seed photons emitted by a quasi-thermal population of relativistic electrons (e.g. Gallo et al.~\cite{g07}; Plotkin et al.~\cite{p15}). In the accretion -- jet scenario which achieved great success in the hard states of BHXTs (Yuan \& Narayan~\cite{yn14}), it is shown that the X-ray emission of the quiescent state of BHXTs will be the optically-thin synchrotron emission from the jet rather than the hot accretion flow (e.g. Yuan \& Cui~\cite{yc05}; Pszota et al.~\cite{p08}; XYM14) . This prediction is supported by several observations (see XYM14 for a recent summary). (1), the quiescent-state spectral energy distribution (SED) of individual sources can be well modelled by the jet theory (Pszota et al.~\cite{p08}; XYM14); (2), deep {\it XMM-Newton} X-ray observations with high signal-to-noise (S/N) ratio, on sources V404 Cyg, GRO J1655--40 and XTE J1550--564 (less confirmative due to relatively poorer S/N ratio), indicate that their X-ray spectra are precisely power-law, without any curvatures (Pszota et al.~\cite{p08}); (3), statistically, the value of X-ray photon index of the quiescent state is roughly a constant, independent of the X-ray luminosity (Plotkin, Gallo \& Jonker~\cite{p13}; Yang et al.~\cite{Yang15}). In addition, through fitting the SEDs in extremely low-luminosity active galactic nuclei (LLAGNs), it is also shown that the X-ray spectrum of those `quiescent' AGNs could be well fitted by the jet model (e.g. Wu, Yuan \& Cao~\cite{wu07}; Yuan, Yu \& Ho ~\cite{y09}; Yu, Yuan \& Ho~\cite{yu11}).

We in this work aim to provide additional support to the jet scenario of the quiescent state of BHXTs. We first in Sec.\ \ref{sec:obs} give the backgrounds and observations of the two sources investigated here, A0620--00 and XTE J1118+480, with a focus on the quasi-simultaneous multi-band (from radio to X-ray) observations in their quiescent states. We note that both sources have been modelled under the maximally efficient jet model (A0620--00: Gallo et al.~\cite{g07}. XTE J1118+480: Plotkin et al.~\cite{p15}). We subsequently in Sec.\ \ref{sec:accmodel} describe the accretion -- jet model we used, and then in Sec.\ \ref{specfit} provide a comprehensive multi-band spectral fitting of the spectra. The last section is devoted to discussions and a brief summary.

\section{Observations of quiescent states of BHXTs}\label{sec:obs}
\subsection{Observations of A0620--00 in quiescent state}\label{sec:a0620}

A0620--00, discovered in 1975, is a low-mass X-ray binary which has been in quiescent state for almost 40 years. As summarized in Table\ \ref{tab1}, it locates at a distance $d=1.06\pm0.12\ {\rm kpc}$ (Cantrell et al.~\cite{Cantrell10}). The orbital period is $P_{\rm orb} = 7.75\ {\rm hr}$ (Johannsen et al.~\cite{j09}) and the line-of-sight inclination angle of the binary system is also constrained to be $\theta=51.0\degr\pm0.9\degr$. The mass and the spin of the black hole are, respectively, $\mbh=6.6\pm0.25\ \msun$ (Cantrell et al.~\cite{Cantrell10}), and $a_\ast=0.12\pm0.19$ (Gou et al.~\cite{Gou10}). The mass and size of the companion star are respectively $0.40\ \msun$ and $0.56\ R_\odot$. Besides, the companion has a spectral type K5V with a temperature of 4400 K (Cantrell et al.~\cite{Cantrell10}).

To our knowledge, there are two broadband near-simultaneous observations on this source, at X-ray luminosity $L_{\rm X} \sim 10^{-8.5}\ \ledd$. The first is in August, 2005 (Gallo et al.~\cite{g07}), where they emphasized on the {\it Spitzer} observation. Later on March 23-25, 2010, Froning et al.~(\cite{Froning11}) carried out contemporaneous X-ray (by {\it Swift}/XRT), ultraviolet (by HST/COS, HST/STIS, and {\it Swift}/UVOT), optical (by {\it Swift}/UVOT and SMARTS/ANDICAM), near-infrared (by Keck and SMARTS/ANDICAM), and radio (by ATCA) observations. For the details of both instruments involved and data reduction, the readers are referred to Froning et al.~(\cite{Froning11}). We also compile the infrared {\it WISE} observation of this source, which is observed on March 19, 2010, i.e. only one week earlier (Wang \& Wang~\cite{wang14}). The observational data is shown in the left panel of Fig.\ \ref{fig:sed}. We have two notes here. First, there are too few photons detected by {\it Swift}/XRT to constrain the X-ray photon index. We can only adopt $\Gamma$ from previous sensitive observations, by e.g. {\it Chandra}, with similar X-ray fluxes (See Froning et al.~\cite{Froning11} for more details). Second, the ATCA radio observations at 5.5 GHz and 9 GHz are only upper limits. Again, we take the data from Gallo et al.~(\cite{g07}), which has similar IR and X-ray fluxes to the new observations, as compensatory data.

\begin{table}[h]
\begin{center}
%\vspace{0.1 cm}
\caption[]{Basic properties of individual sources.}\label{tab1}\tiny	
 \begin{tabular}{cccccccc}
  \hline\noalign{\smallskip}
Sources &  Distance   & Black hole mass  & Orbital period  & Inclination & Binary separation & Companion star & References \\
          &  $d$ (kpc)   &   $\mbh\ (\msun)$ & $P_{\rm orb}$ (hr) & $\theta$ (degree) & $a$ (cm) & (MK Type, temperature, radius)& \\
\hline
A0620--00    &  $1.06\pm0.12$   &  $6.6\pm0.25$ & $7.75$ & $51.0\pm0.9$ & $2.6\times 10^{11}$  & K5V, 4400 K, $0.50-0.56\ R_\odot$ & J09, C10 \\
XTE J1118+480    &  $1.72\pm0.10$  & $7.5\pm0.6$ & $4.08$ & $68-79$ & $1.8\times10^{11}$  & K7V--M1V, 4000 K, $0.6\ R_\odot$ & T04, G06, K13\\
  \noalign{\smallskip}\hline																					
\end{tabular}
\end{center}	
\footnotesize{References: T04 -- Torres et al.~(\cite{t04}); G06 -- Gelino et al. (\cite{g06}); J09 -- Johannsen et al.~(\cite{j09}); C10 -- Cantrell et al.~(\cite{Cantrell10}); K13 -- Khargharia et al.~(\cite{Khargharia13}).}							
\end{table}

\subsection{Observations of XTE J1118+480 in quiescent state}

XTE J1118+480, discovered by the {\it Rossi X-Ray Timing Explorer} ({\it RXTE}) all-sky monitor on March 29, 2000, is a low-mass X-ray binary that undergone several outbursts. As summarized in Table\ \ref{tab1}, it locates at a distance $d=1.72\pm0.10\ {\rm kpc}$ (Gelino et al.~\cite{g06}), with a black hole mass $\mbh=7.5\pm0.6\ \msun$ (Khargharia et al.~\cite{Khargharia13}). The orbital period is $P_{\rm orb} = 4.08\ {\rm hr}$ (Torres et al.~\cite{t04}) and the line-of-sight inclination angle of the binary system is also constrained to be $\theta=68\degr - 79\degr$ (Khargharia et al.~\cite{Khargharia13}). The mass and size of the companion star are respectively $0.3\pm0.2\ \msun$ (Mirabel et al.~\cite{mirabel01}) and $0.6\ R_\odot$. Besides, the companion has a spectral type K7V--M1V with a temperature of 4000 K (Khargharia et al.~\cite{Khargharia13}).

Broadband near-simultaneous observations of XTE J1118+480 are carried out on June 27-28, 2013 (Gallo et al.~\cite{g14}; Plotkin et al.~\cite{p15}). Waveband included in these observations are X-ray (by {\it Chandra}/ACIS), ultraviolet (by {\it Swift}/UVOT), optical (by WHT/ACAM), near-infrared (by WHT/LIRIS and 2MASS), and radio (by VLA). For the details of both instruments involved and data reduction, the readers are referred to Plotkin et al. (\cite{p15}). Note that we also have mid-infrared (by {\it WISE}) observations of this source in quiescent state, on March 10, 2010 (Wang \& Wang~\cite{wang14}). The observational data of XTE J1118+480 quiescence is shown in the right panel of Fig.\ \ref{fig:sed}.

\section{the accretion -- jet model}\label{sec:accmodel}

In this work, we take the accretion -- jet model, which has been successfully applied to the BHXTs in their hard states and the LLAGNs (see Yuan \& Narayan~\cite{yn14} for a recent review). In this model, three components are considered (YCN05), i.e. an outer (possibly irradiated) SSD, which has an outer boundary $R_{\rm out, SSD}$ and will be truncated at radius $R_{\rm tr}$, an inner hot accretion flow within $R_{\rm tr}$, and finally a collimated relativistic jet perpendicular to plane of the accretion flow. From observational point of view, the three components of the accretion -- jet model are prominent in different wavebands. Very roughly, the truncated SSD, the hot accretion flow and the jet, respectively, contribute mainly to the radiation in the infrared--ultraviolet, X-ray and radio bands.

We in the subsequent subsection provide more detailed information about the accretion-jet model, especially the jet component, since we try to advocate that most of the X-ray radiation is the synchrotron emission from the jet (see Yuan \& Cui~\cite{yc05}; Pszota et al.~\cite{p08}; XYM14).

\subsection{Hot accretion flow model}\label{sec:adafmodel}

The hot accretion flow adopted in our model is an advection-dominated accretion flow (ADAF; Narayan \& Yi~\cite{ny94}). Compared with the ADAF model adopted in Narayan, McClintock \& Yi~(\cite{n96}), two important progresses are taken into account (see Xie \& Yuan~\cite{xy12}; Yuan \& Narayan~\cite{yn14} for summaries), i.e. the existence of outflow and the viscous/turbulent heating onto electrons. For the outflow, both the observational evidences and theoretical simulations find the strong wind exists in the hot accretion flow (Yuan, Bu \& Wu~\cite{ybw12}; Yuan \& Narayan~\cite{yn14}; Yuan et al.~\cite{y15}). Consequently the accretion rate follows $\dot{M}(R)\propto R^s$, where index $s$ is the outflow parameter, ranging from $\sim0.4$ to $\sim 0.8$ (Yuan, Wu \& Bu~\cite{ywb12}).

The fraction of turbulent viscous/turbulent heating rate that goes into electrons is devoted as $\delta$. The microphysics of viscous/turbulent heating is still under active research and numerous processes are proposed, i.e. MHD turbulence (Quataert~\cite{q98}; Blackman~\cite{b99}; Lehe et al.~\cite{lehe09}), magnetic reconnection (Bisnovatyi-Kogan \& Lovelace~\cite{bl97}; Quataert \& Gruzinov~\cite{q99}; Ding et al.~\cite{ding10}), or dissipation of pressure anisotropy (Sharma et al.~\cite{Sharma07}; Sironi \& Narayan~\cite{sn15}). Different microphysics will result in different value of $\delta$, and it is likely in the range from 0.1 to 0.5 (see Yuan \& Narayan~\cite{yn14}). Throughout this work, we adopt $\delta=0.1$ in our numerical calculations (ref. Yang et al.~\cite{Yang15}).

\subsection{Jet model}\label{sec:jetmodel}

Our jet model is phenomenological (see YCN05, Xie \& Yuan~\cite{xy15} for more details). The composition is assumed to be normal plasma, i.e. electrons and protons. The bulk Lorentz factor of the compact jet is set to $\Gamma_{\rm jet} =1.2$, a typical value for the jets in the hard state of BHXTs (Gallo, Fender \& Pooley ~\cite{g03}; Fender~\cite{f06}). The opening angle is $\theta_{\rm jet} = 0.1$ (or equivalently $\theta_{\rm jet}\sim5.7\degr$).

Within the jet itself, different shells of the moving plasma are assumed to have different velocities. When the faster but later shells catch up with the slower and earlier ones, internal shocks occur. The spatial filling factor of the internal shock shells is assumed to be $0.1$. In these internal shocks a fraction ($\xi$) of the electrons will be accelerated into a power-law energy distribution, with the index of $p_{\rm e}$. Due to the strong radiative cooling, the high-energy part (self-consistently determined in our calculations) of the accelerated power-law electrons will be cooled down, and their distribution index will be $p_{\rm e}+1$ (Rybicki \& Lightman~\cite{rl79}). Two additional parameters, $\epsilon_e$ and $\epsilon_{\rm _B}$, are also included to quantify the fraction of the shock energy that goes into electrons and magnetic fields, respectively. For simplicity, all these microphysical parameters are assumed to be constant along the jet direction.

With above parameters, we can calculate the synchrotron emission from these accelerated power-law electrons. The parameter dependence is extensively discussed in XYM14. The spectral energy distribution (SED) of the jet (or more generally the power-law electrons) in general is fairly simple. The high energy part (e.g. UV and X-ray bands) is power-law, with photon index $\Gamma \approx 1+(p_{\rm e}+1-1)/2 = 1+p_{\rm e}/2$ (Rybicki \& Lightman~\cite{rl79}). The spectrum of low energy part (e.g. radio up to IR) is also power-law, but is flat or slightly inverted because of self-absorption.

\begin{figure}
\centering
\includegraphics[width=7.cm]{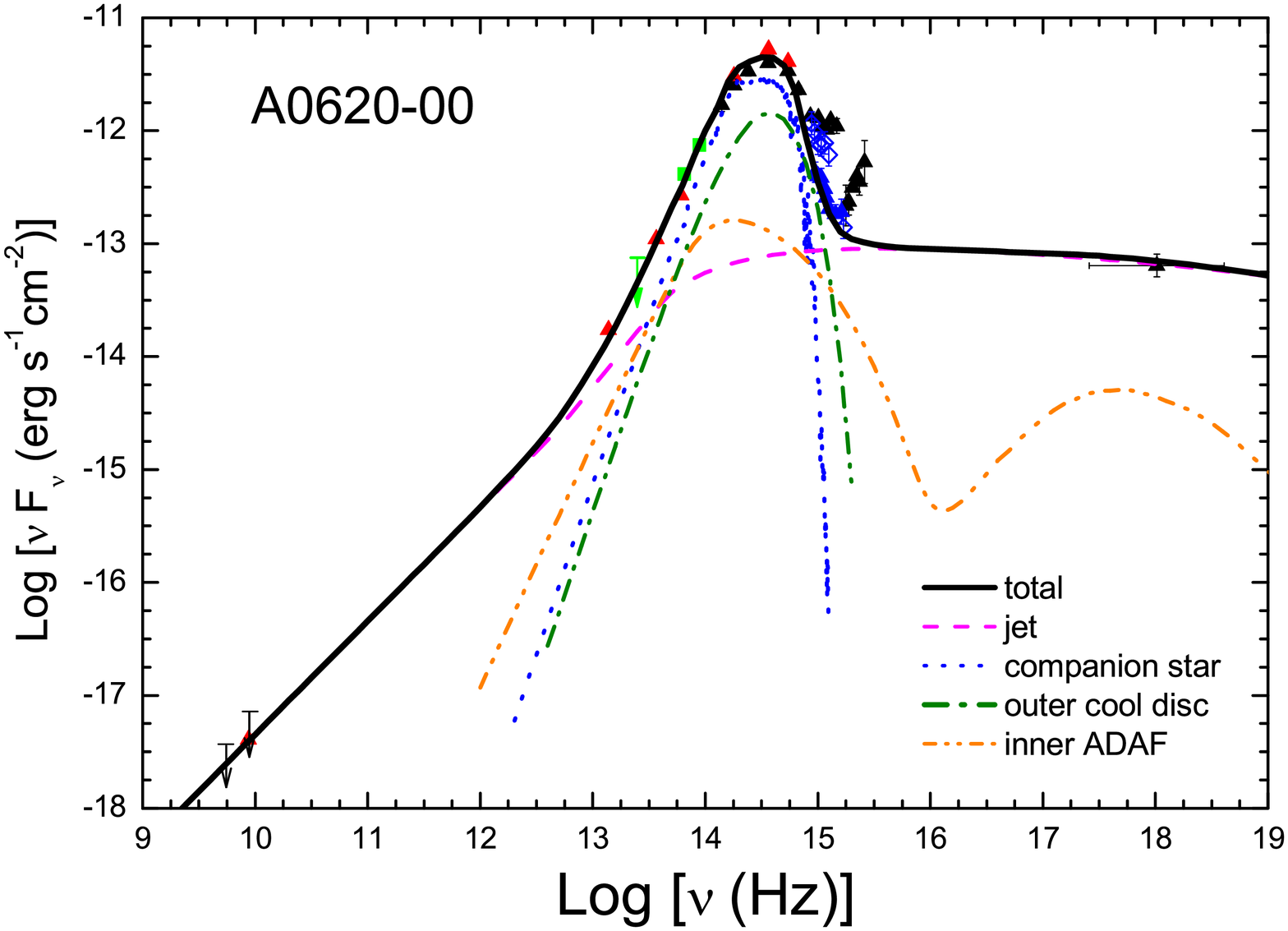}
\includegraphics[width=7.cm]{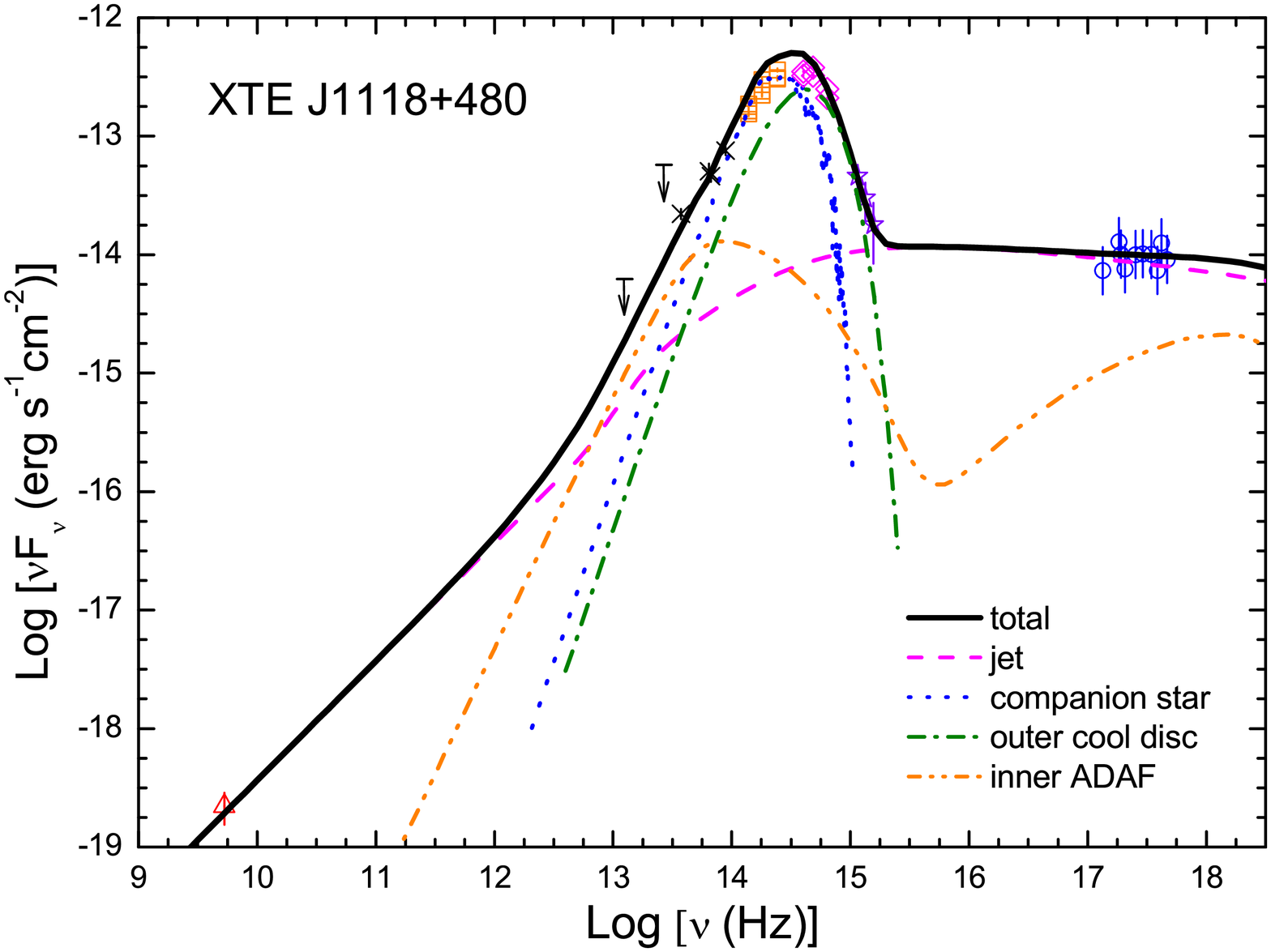}
\caption{Broadband spectral energy distribution (SED) and the spectral fitting results (curves labelled) of two BHXBs (A0620--00 \& XTE J1118+480) in their quiescent states. {\it Left panel}: A0620--00. The solid black triangles are from near-simultaneous observation by Froning et al.~(\cite{Froning11}), while the solid red triangles and green squares show the non-simultaneous data from Gallo et al.~(\cite{g07}) and Wang \& Wang (\cite{wang14}), respectively. {\it Right panel}: XTE J1118+480. The observational data are taken from Plotkin et al. (\cite{p15}). Symbols are for the radio (triangle), (non-simultaneous) IR (arrows for upper limits; crosses for detections), NIR and optical (squares), UV (stars), and X-ray (circles) data points.}
\label{fig:sed}
\end{figure}

\section{Results}\label{specfit}
Before presenting our results, we first comment on the ADAF model for the quiescent state (e.g., Narayan, McClintock \& Yi~\cite{n96}; Narayan, Barret \& McClintock~\cite{n97}), where the X-ray radiation is believed to be emitted from ADAF. Observationally, the ADAF model suffers two problems. First, once the X-ray flux is fitted, the UV (with $\nu\ga 10^{15}\ {\rm Hz}$) emission predicted by ADAF model is substantially higher compared to observations (ref. Fig 1a in Yuan \& Narayan~\cite{yn14} for the spectrum of ADAF at low accretion rate). This argument is adopted by Hynes et al.~\cite{h09} for the observation of V404 Cyg in its quiescence, where they found that the ADAF model produces too strong UV emission, by a factor of $\sim 10$, for a given X-ray flux. Second, the radio observation of quiescent state indicates the existence of jet, or, more generally non-thermal electrons. These non-thermal electrons will also likely to emit in the X-ray band, i.e. the jet contribution in the X-ray band should be taken into account. Because of the above two reasons, the ADAF model seems unlikely.

We thus apply the accretion -- jet model to fit the quasi-simultaneous multi-band observations of both A0620--00 and XTE J1118+480. Below we detail our methods and results.

{\bf We first take into account the emission from the companion star.} We adopt the stellar atmosphere model by Kurucz (\cite{k93}). As shown by the blue dotted curves in Fig.\ \ref{fig:sed}, the companion dominates the radiation in mid-IR-optical bands.

{\bf We secondly constrain the jet emission contribution.} Observationally, the X-ray emission of the quiescent state can be fitted by a power-law, i.e. with photon index to be $\Gamma=2.26\pm0.18$ (McClintock et al.~\cite{mcc03}; it has similar X-ray flux as this observation) for A0620--00 and $\Gamma=2.02\pm0.41$ (Plotkin et al.~\cite{p15}) for XTE J1118+480. Extending the power-law spectrum to lower energy band (e.g. UV with $\nu\sim10^{15}$ Hz), we find that it can naturally compensate the excess emission in UV. For the value of $\Gamma$ of both sources, we pre-fix in our jet model $p_{\rm jet}=2.3$, a value consistent with the diffusive shock acceleration theory. There are four free parameters left, $\mdot_{\rm jet}, ~\xi, ~\epsilon_e$ and $\epsilon_{\rm _B}$ (see XYM14 for discussions on the effects of these parameters). Through spectral modelling (i.e. fluxes in radio, IR, UV and X-ray bands), we can obtain the fitting physical parameters (list in Table. \ref{tab2}), which are all within the typical range obtained in GRB afterglow (YCN05). We find that the radiation in mid-UV up to X-ray bands is the optically thin part of the synchrotron radiation. On the other hand, the radiation in radio up to far-IR bands is the optically thick self-absorbed synchrotron radiation, and the spectrum is flat with spectral index $\alpha_{R-IR} \approx 0$.

{\bf Thirdly, we check the radiative contribution from the inner hot accretion flow and the outer cold disc.} Generally the radiation from the outer cold disc is difficult to constrain in quiescent states, due to the dominance of the companion star, e.g. the case in V404 Cyg (XYM14). However, the emission of the bluer-optical band up to the UV band of A0620--00 and XTE J1118+480 in their quiescent states show clear flux excess compared to that from the companion (cf. Fig.\ \ref{fig:sed}). Such excess can be understood naturally as emission from the outer cold disc. We considered the irradiated cold disc model, with three model parameters, i.e. the accretion rate ($\mdot_{\rm SSD}$), and the outer and inner radii ($R_{\rm out, SSD}$ and $R_{\rm in, SSD}$, note that $R_{\rm in, SSD} \equiv R_{\rm tr}$). We emphasis that the SED coverage in the UV band is of crucial importance to determine the transition radius $R_{\rm tr}$. Emission from the outer irradiated cold disc is shown as green dot-dashed curves in Fig.\ \ref{fig:sed}.

Subsequently we assume that the mass-loss rate in the jet ($\mdot_{\rm jet}$) at $5\ R_{\rm s}$ (here $R_{\rm s}$ is the Schwarzschild radius of the black hole) is $\sim 6\%$ of the accretion rate at 5 $R_{\rm s}$, a reasonable value in the coupled accretion -- jet model (see e.g. Fig.\ 2 in Yuan \& Cui~\cite{yc05}). With the accretion rates at $R_{\rm tr}$ and $5\ R_{\rm s}$, the outflow parameter $s$ can be constrained then, i.e. $s \approx 0.4$ for A0620-00 and $s \approx 0.6$ for XTE J1118+480. Both values are consistent with the suggested range ($s \sim 0.4-0.8$) from recent large-scale numerical simulations of hot accretion flows (e.g. Yuan, Wu \& Bu~\cite{ywb12}). Other basic parameters of the hot accretion flow include the viscous parameter $\alpha_{\rm vis}=0.3$, the magnetic parameter (defined as the ratio of the gas to the magnetic pressure) $\beta=9$, and the fraction of viscous heating onto electrons $\delta=0.1$ (ref. Yang et al.~\cite{Yang15}). Emission from the inner ADAF is shown as dot-dot-dashed curves in Fig.\ \ref{fig:sed}, where the lower-$\nu$ peak is the synchrotron and the higher-$\nu$ peak is bremsstrahlung (dominated) with a negligible fraction of inverse Compton. Evidently from Fig.\ \ref{fig:sed}, emission from ADAF is not important at any wavebands for the quiescent state of BHXTs (see also XYM14).

We summarize the detailed modelling parameters of the coupled accretion -- jet model for the quiescent state of A0620--00 and XTE J1118+480, and list them in Table\ \ref{tab2}. For comparison and completeness, we also list the fitting parameters of V404 Cyg in its quiescent state (ref. XYM14). Obviously the fitting parameters from these three BHXTs are quite similar to each other.

We note that independent estimation and/or measurement on both $R_{\rm tr}$ and $\mdot_{\rm SSD}$ has been developed. For example, the transition radius $R_{\rm tr}$ can be estimated from the low frequency (roughly at mHz level) quasi-periodic-oscillation (QPO). For XTE J1118+480, Shahbaz et al. (\cite{Shahbaz05}) carried out  high time-resolution multi-colour ULTRACAM observation on this source at its quiescent state, which has similar X-ray flux to the observations adopted here, and found that the power density spectrum of the light curves could be described by a power-law model plus a broad QPO with frequency $\sim2$ mHz. Assuming it is the Keplerian rotational frequency at transition radius (Giannios \& Spruit~\cite{g04}), they estimated $R_{\rm tr}\approx 8000\ R_{\rm s}$, consistent with the value derived from spectral fitting.

Besides, Lasota (\cite{l00}) provided a formulae to estimate mass accretion rate at large radius for systems in their quiescent state, i.e. $\mdot(R) \approx 4.0\times10^{15} \left(M_{\rm BH}/1\ \msun\right)^{-0.88}\ \left(R/10^{10}\ {\rm cm}\right)^{2.65}\ {\rm g\ s^{-1}}$. Applying the formulae to A0620-00 and XTE J1118+480, we find the accretion rates of the outer cold disc are, respectively, $\approx5\times10^{-4}\ \mdot_{\rm Edd}$ and $\approx3\times10^{-4}\ \mdot_{\rm Edd}$, where $\dot{M}_{\rm Edd} = 10\ L_{\rm Edd}/c^2$ is the Eddington accretion rate. Again, these values are consistent, within a factor of $\sim 2$, with the values from spectral modelling results (cf. Table \ref{tab2}).

\begin{table}[h]
\begin{center}
\vspace{0.1 cm}
\caption[]{Modelling parameters of the coupled accretion -- jet model}\label{tab2}	
%\vspace{0.2 cm}	
\begin{tabular}{c c c c c | c c c c c}	
\hline\noalign{\smallskip}
Sources &  $\mdot_{\rm SSD}$ & $R_{\rm out, SSD}$ & $R_{\rm tr}$   & $s$  &$\mdot_{\rm jet}$ & $p_{\rm e}$ &$\epsilon_{\rm e}$ &$\epsilon_{\rm _B}$ & $\xi$  \\
 &  ($10^{-4}\ \medd$) &  ($10^4\ R_{\rm s}$) & ($10^4\ R_{\rm s}$) &   &  ($10^{-6}\ \medd$)  & & & \\
\hline
A0620--00 &$4.$ &2.5&1.&0.4&$1.2$&2.3&0.01&0.04&0.06\\
XTE J1118+480&$6.$&2.&0.8&0.6&$0.43$&2.3&0.025&0.05&0.04 \\
V404 Cyg &$50.$&5.&2.&0.6&$3.$&2.4&0.04&0.03&0.08\\
  \noalign{\smallskip}\hline																					
\end{tabular}
\end{center}
\end{table}

\section{Summary and Discussion}\label{summary}
%\subsection{strong evidences in favor of the jet-dominated model}

The radiative properties of the quiescent state, either in BHXTs or in normal galaxies, remain unclear, and numerous efforts are devoted in this field. Thanks to the high-sensitivity X-ray telescopes and the joint multi-band (quasi-)simultaneous observations, together with long-term narrow-band monitoring, we confirm previous theoretical results (see XYM14 for summaries) that the emission of the quiescent states of BHXTs is dominated by the radiation from the compact relativistic jet. The outer thin disc and the inner hot accretion flow generally play negligible roles in radiation, while the companion dominates the emission between mid-IR and optical bands. We also illustrate that the jet-dominated quiescent state model can explain most of the observational features (see XYM14 for details). Below we provide several discussions of our results.

\subsection{The UV spectrum of A6020--00}

Froning et al. (\cite{Froning11}) in their {\it Hubble} observations, include both the near-UV (NUV; $\nu<1.5\times10^{15}$ Hz) STIS instrument and far-UV (FUV; $\nu>1.5\times10^{15}$ Hz) COS instrument. However, clear discrepancy between these two instruments are observed (see left panel in Fig.\ \ref{fig:sed}), i.e. the FUV spectrum is rather blue (flux $F_\nu\propto \nu^{-\alpha}$, with $\alpha>-1$.). Besides, the NUV spectrum, which is also moderately blue, is also different from previous observation.

The UV spectrum is discussed extensively in Froning et al. (\cite{Froning11}). More directly, they fitted this part with a new thermal component, and found it moderately hot and compact, with temperature $\sim 10^4$ K and size $\sim 4\times10^9$ cm. Several possibilities are proposed, most likely it is the bright spot (the accretion stream-disk impact point). It can also be related to the transition zone (close to the transition radius) between the outer cold disc and the inner hot accretion flow. In the current work, we do not include this complexity.

\subsection{Radio/X-ray correlation in quiescent state}

One important but unclear question is the correlation slope between the radio luminosity and X-ray luminosity, in the form $L_R\propto L_X^p$, where $p$ is the correlation slope, in quiescent state of black hole sources. Tight and strong radio/X-ray correlation is observed in both AGNs and BHXTs (e.g. Merloni et al.~\cite{m03}; Corbel et al.~\cite{c13}), with $p\approx0.6$. Under the accretion -- jet model, Yuan \& Cui (\cite{yc05}) theoretically predict that the correlation will steepens, with $p\sim 1.23$, when the X-ray luminosity $L_X$ is below a critical value, $L_{X, crit}\sim 10^{-5}-10^{-6}~\ledd$ ($\ledd$ is the Eddington luminosity), where the X-ray emission will come from the jet rather than the hot accretion flow normally observed in the hard/bright states (Yuan \& Cui~\cite{yc05}). This prediction is confirmed by later works, i.e., data from all available LLAGNs satisfying $L_X\la L_{X, crit}$ follows $L_R\propto L_X^{1.22}$ (Pellegrini et al.~\cite{p07}; Wu, Yuan \& Cao~\cite{wu07}; Wrobel et al.~\cite{w08}; Yuan, Yu \& Ho~\cite{y09}; de Gasperin et al.~\cite{de11}; Younes et al.~\cite{younes12}).

In the case of BHXTs, however, the answer is not so clear. We here emphasize two cautions. First, although the observations of A0620--00 and XTE J1118+480 can be fitted well within the jet model for combined data set (cf. \S.\ \ref{specfit}), the radio data is still not very good/robust (see also Yuan \& Narayan~\cite{yn14}). For example, the most recent radio detection of XTE J1118+480 in its quiescent state is marginal, with only at 3$\sigma$ level (Gallo et al.~\cite{g14}), while A0620--00 is under detection (Froning et al.~\cite{Froning11}) or also at 3$\sigma$ level (Gallo et al.\cite{g07}).

Second, there might be systematical differences between low-luminosity AGNs and BHXTs, in the sense of different mass supply. The BHXTs accrete material from its companion, where the binary separation will constrain the size and location of the outer cold disc (cf. Table \ref{tab1}). Consequently the jet properties ($\Gamma_{\rm jet}, \xi, \epsilon_e, \epsilon_{\rm _B}$) and their relationship with the decreasing mass loss rate into the jet $\mdot_{\rm jet}$ in BHXTs and AGNs may be different. One direct evidence is that, the bulk Lorentz factor of the compact jet in BHXTs is $\Gamma_{\rm jet}\lesssim 1.6$ (e.g. Fender, Belloni \& Gallo~\cite{f04}), while it is likely much higher in AGNs, with typical value $\Gamma_{\rm jet}\sim 10$. Besides, the magnetic field configuration in BHXTs may also be different from that in AGNs, which may eventually also affect the jet formation/acceleration processes. Further efforts, both theoretical and observational, are still urged to understand the physics of jet.

\normalem
\begin{acknowledgements}
QXY is grateful to the anonymous referee for constructive suggestions, and Prof. Feng Yuan, Dr. Fu-Guo Xie and Yaping Li for useful discussions. This work was supported by the National Basic Research Program of China (973 Program, grant 2014CB845800), the Natural Science Foundation of China (grants 11203057, 11103061, 11133005 and 11121062), and the Strategic Priority Research Program ``The Emergence of Cosmological Structures" of the Chinese Academy of Sciences (Grant XDB09000000).
\end{acknowledgements}


\begin{thebibliography}{99}

\bibitem[2010]{b10} Belloni T.~M., 2010, in ``The Jet Paradigm - From Microquasars to Quasars'', ed. T. Belloni,  Lecture Notes in Physics, Springer-Verlag, Berlin, 794, 53

\bibitem[2014]{bc14} Bernardini F., Cackett E. M., 2014, \mnras, 439, 2771

\bibitem[1997]{bl97} Bisnovatyi-Kogan G. S., Lovelace R. V. E., 1997, \apj, 486, L43

\bibitem[1999]{b99} Blackman E. G., 1999, \mnras, 302, 723

\bibitem[2010]{Cantrell10} Cantrell A.~G., et al., 2010, \apj, 710, 1127

\bibitem[2013]{c13} Corbel S., Coriat M., Brocksopp C., Tzioumis A.~K., et al., 2013, \mnras, 428, 2500

\bibitem[2006]{c06} Corbel S., Tomsick J.~A., Kaaret P., 2006, \apj, 636, 971

\bibitem[2011]{de11} de Gasperin F., Merloni A., Sell P., et al., 2011, \mnras, 415,2910

\bibitem[2010]{ding10} Ding J., Yuan F., Liang E., 2010, \apj, 708, 1545

\bibitem[2007]{done07} Done C., Gierlinski, M., Kubota, A., 2007, A\&ARv, 15, 1

\bibitem[1997]{esin97} Esin A.~A., McClintock J.~E., Narayan R., 1997, \apj, 489, 865

\bibitem[2006]{f06} Fender R.~P., 2006, in ``Compact stellar X-ray sources'', Eds. W. Lewin \& M. van der Klis, Cambridge Astrophysics Series, No. 39. Cambridge, UK: Cambridge University Press, 381

\bibitem[2004]{f04} Fender R.~P., Belloni T.~M., Gallo E., 2004, \mnras, 355, 1105

\bibitem[2006]{Ferr06} Ferreira J., Petrucci P. O., Henri G., Saug\'{e} L., Pelletier G., 2006, A\&A, 447, 813

\bibitem[2011]{Froning11} Froning C.~S., et al., 2011, \apj, 743, 26

\bibitem[2003]{g03} Gallo E., Fender R.~P., Pooley G.~G., 2003, \mnras, 344, 60

\bibitem[2007]{g07} Gallo E., Migliari S., Markoff S., Tomsick J.~A. et al., 2007, \apj, 670, 600

\bibitem[2014]{g14} Gallo E., et al., 2014, \mnras, 445, 290

\bibitem[2006]{g06} Gelino D. M., Balman S., K{\i}z{\i}lo{\v g}lu {\"U} ., Y{\i}lmaz A., Kalemci E., Tomsick J. A., 2006, \apj, 642, 438

\bibitem[2004]{g04} Giannios D., Spruit H. C., 2004, A\&A, 427, 251

\bibitem[2010]{Gou10} Gou L., McClintock J.~E., Steiner J.~F., Narayan R., Cantrell A.~G., Bailyn C.~D., Orosz J.~A., 2010, \apj, 718, L122

\bibitem[2008]{ho08} Ho L.~C., 2008, \araa, 46, 475

\bibitem[2009]{ho09} Ho L.~C., 2009, \apj, 699, 626

\bibitem[2005]{hb05} Homan, J., Belloni, T., 2005, Ap\&SS, 300, 107

\bibitem[2009]{h09} Hynes R.~I., Bradley C.~K., Rupen M., Gallo E. et al., 2009, \mnras, 399, 2239

\bibitem[2004]{h04} Hynes R.~I., Charles P.~A., Garcia M.~R., Robinson E.~L. et al., 2004, \apj, 611, L125

\bibitem[2009]{j09} Johannsen, T., Psaltis, D., \& McClintock, J. E. 2009, \apj, 691, 997

\bibitem[2013]{Khargharia13} Khargharia J., Froning C.~S., Robinson E.~L., Gelino D.~M., 2013, \aj, 145, 21

\bibitem[2002]{Kong02} Kong A.~K.~H., McClintock J.~E., Garcia M.~R., Murray S.~S., Barret D., 2002, \apj, 570, 277

\bibitem[1993]{k93} Kurucz R.~L., 1993, SYNTHE Spectrum Synthesis Programs and Line Data (CD-ROM), Smithsonian Astrophysical Observatory, Cambridge, MA

\bibitem[2008]{k08} Kylafis N.~D., Papadakis I.~E., Reig P., Giannios D., Pooley G.~G., 2008, A\&A, 489, 481

\bibitem[2000]{l00} Lasota J.~P., 2000, A\&A, 360, 575

\bibitem[2009]{lehe09} Lehe R., Parrish I. J., Quataert E., 2009, \apj, 707, 404

\bibitem[2002]{liu02} Liu B.~F., Mineshige S., Meyer F., Meyer-Hofmeister E., Kawaguchi T., 2002, \apj, 575, 117

\bibitem[2007]{liu07} Liu B. F., Taam R. E., Meyer-Hofmeister E., Meyer F., 2007, \apj, 671, 695

\bibitem[2005]{m05} Markoff S., Nowak M.~A., Wilms J., 2005, \apj, 635, 1203

\bibitem[2003]{mcc03} McClintock J.~E., Narayan R., Garcia M.~R., Orosz J.~A., et al., 2003, \apj, 593, 435

\bibitem[2003]{m03} Merloni A., Heinz S., di Matteo T., 2003, \mnras, 345, 1057

\bibitem[2001]{mirabel01} Mirabel I. F., Dhawan V., Mignani R. P., Rodrigues I., Guglielmetti F., 2001, Nature, 413, 139

\bibitem[1997]{n97} Narayan R., Barret D., McClintock J.~E., 1997, \apj, 482, 448

\bibitem[2002]{n02} Narayan R., Garcia M.~R.,  McClintock J.~E., 2002, in The Ninth Marcel Grossmann Meeting, eds. V.~G. Gurzadyan, R.~T. Jantzen, R. Ruffini (Singapore: World Scientific), 405

\bibitem[2008]{nm08} Narayan R., McClintock J.~E., 2008, NewAR, 51, 733

\bibitem[1996]{n96} Narayan R., McClintock, J.~E., Yi, I. 1996, \apj, 457, 821

\bibitem[1994]{ny94} Narayan R., Yi I., 1994, \apj, 428, L13

\bibitem[2010]{pel10} Pellegrini S., 2010, \apj, 717, 640

\bibitem[2007]{p07} Pellegrini S., Siemiginowska A., Fabbiano G., et al., 2007, \apj,667, 749

\bibitem[2013]{p13} Plotkin R.~M., Gallo E., Jonker P.~G., 2013, \apj, 773, 59

\bibitem[2015]{p15} Plotkin R.~M., Gallo E., Markoff S., Homan J., Jonker P.~G., Miller-Jones J.~C.~A., Russell D.~M., Drappeau S., 2015, \mnras, 446, 4098

\bibitem[2008]{p08} Pszota G., Zhang H., Yuan F., Cui W., 2008, \mnras, 389, 423

\bibitem[2013]{ql13} Qiao E. L., Liu B. F., 2013, \apj, 764, 2

\bibitem[1998]{q98} Quataert E., 1998, \apj, 500, 978

\bibitem[1999]{q99} Quataert E., Gruzinov A., 1999, \apj, 520, 248

\bibitem[2015]{r15} Rana V., et al., 2015, arXiv: 1507.04049

\bibitem[2006]{rm06} Remillard R.~A., McClintock J.~E., 2006, \araa, 44, 49

\bibitem[2011]{r11} Reynolds M.~T., Miller J.~M., 2011, \apj, 734, L17

\bibitem[2014]{r14} Reynolds M.~T., Reis R.~C., Miller J.~M., Cackett E.~M., Degenaar N., 2014, \mnras, 441, 3656

\bibitem[1979]{rl79} Rybicki G.~B., Lightman A.~P., 1979, Radiative Processes in Astrophysics (New York: Wiley)

\bibitem[2005]{Shahbaz05}Shahbaz T., Dhillon V. S., Marsh T. R., Casares J., Zurita C., Charles P. A., Haswell C. A., Hynes R. I., 2005, \mnras, 362, 975

\bibitem[1973]{ss73} Shakura N.~I., Sunyaev R.~A., 1973, A\&A, 24, 337

\bibitem[2007]{Sharma07} Sharma P., Quataert E., Hammett G. W., Stone J. M., 2007, \apj, 667, 714

\bibitem[2015]{sn15} Sironi L., Narayan R., 2015, \apj, 800, 88

\bibitem[2004]{t04} Torres M. A. P., Callanan P. J., Garcia M. R., Zhao P., Laycock S., Kong A. K. H., 2004, \apj, 612, 1026

\bibitem[2014]{wang14} Wang X., Wang Z., 2014, \apj, 788, 184

\bibitem[2008]{w08}Wrobel J. M., Terashima Y., Ho L. C., 2008, \apj, 675, 1041

\bibitem[2007]{wu07} Wu Q., Yuan F., Cao X., 2007, \apj, 669, 96

\bibitem[2014]{x14} Xie F. G., Yang Q. X., Ma R., 2014, \mnras, 442, L110 (XYM14)

\bibitem[2012]{xy12} Xie F. G., \& Yuan F. 2012, \mnras, 427, 1580

\bibitem[2015]{xy15} Xie F. G., \& Yuan F. 2015, \mnras (submitted, arXiv: 1509.02598)

\bibitem[2015]{Yang15} Yang Q. X., Xie F. G., Yuan F., Zdziarski A. A., Gierli\'{n}ski M., Ho L. C., Yu Z., 2015, \mnras, 447, 1692

\bibitem[2012]{younes12} Younes G., Porquet D., Sabra B., Reeves J. N., Grosso N., 2012, A\&A, 539, 104

\bibitem[2011]{yu11}Yu Z., Yuan F., Ho L. C., 2011, \apj, 726, 87

\bibitem[2012]{ybw12} Yuan F., Bu D., Wu M., 2012, \apj, 761, 130

\bibitem[2005]{ycn05} Yuan F., Cui W., Narayan R., 2005a, \apj, 620, 905 (YCN05)

\bibitem[2005]{yc05} Yuan F., Cui W., 2005b, \apj, 629, 408

\bibitem[2015]{y15} Yuan F., Gan Z., Narayan R., Sadowski A., Bu D., Bai X. N., 2015, \apj, 804, 101

\bibitem[2014]{yn14} Yuan F., Narayan R., 2014, \araa, 52, 529

\bibitem[2012]{ywb12} Yuan F., Wu M., Bu D., 2012, \apj, 761, 129

\bibitem[2009]{y09} Yuan F., Yu Z., Ho L., 2009, \apj, 703, 1034

\bibitem[2004]{zg04} Zdziarski A.~A., Gierlinski M., 2004, Prog. Theo. Phy. Supp., 155, 99

\bibitem[2013]{zx13} Zhang J.~F., Xie F.~G., 2013, \mnras, 435, 1165

\end{thebibliography}
\end{document}